\documentclass[aps,prl,twocolumn,showpacs,a4paper]{revtex4}
\usepackage{graphicx}
\usepackage{amsmath}
\usepackage{hyperref}
\usepackage{units}
\begin{document}
\title{Probing nearest-neighbor correlations of ultracold fermions in an optical
lattice}
\author{Daniel~Greif}
\author{Leticia~Tarruell}\email{tarruell@phys.ethz.ch}
\author{Thomas~Uehlinger}
\author{Robert~J\"{o}rdens}
\author{Tilman~Esslinger} \affiliation{Institute for Quantum Electronics,
ETH Zurich, 8093 Zurich, Switzerland}

\date{\today}
\begin{abstract}
We demonstrate a probe for nearest-neighbor correlations of fermionic
quantum gases in optical lattices. It gives access to spin and density
configurations of adjacent sites and relies on creating additional doubly
occupied sites by perturbative lattice modulation. The measured
correlations for different lattice temperatures are in good agreement with
an \emph{ab initio} calculation without any fitting parameters. This probe
opens new prospects for studying the approach to magnetically ordered
phases.\end{abstract}
 \maketitle

The Fermi-Hubbard Hamiltonian is one of the central models for
understanding strongly correlated electron systems in condensed matter
physics. It incorporates intriguing phenomena such as Mott-insulating
behavior or spin ordered phases and is a prominent candidate for describing
the origin of high $T_{\mathrm{c}}$ superconductivity. Despite tremendous
theoretical effort, several questions still remain open, in particular
concerning the low temperature phases. Here, ultracold atomic gases trapped
in optical lattices offer the opportunity to address these questions in a
very clean way, as they constitute an almost ideal implementation of the
Hubbard model \cite{Bloch2008, Esslinger2010}. The recent realization of a
fermionic Mott insulator \cite{Jordens2008, Schneider2008}
demonstrates the unique tunability and control of these systems.

Considerable experimental efforts are currently directed towards reaching
the low-temperature regime of quantum magnetism in a two-component quantum
gas. Detection of the antiferromagnetically ordered state has been
proposed via noise correlation or Bragg scattering measurements
\cite{Corcovilos2010, Altman2004}. These observables reveal long range
spin ordering and consequently only show a strong signature well below the
critical temperature. However, a probe sensitive to local magnetic
correlations \cite{Trotzky2010,Baur2010} for studying the approach to
magnetic ordering of fermions close to the transition point is missing so
far.

In this Letter we demonstrate a simple method for probing the
nearest-neighbor correlations of strongly interacting repulsive fermionic
gases in optical lattices. The correlation function is given by
\begin{equation}
    \mathcal{P}_{i,i+1}=\sum_{\sigma}
    \langle n_{i,\sigma} n_{i+1,\bar{\sigma}}
    \left(1-n_{i,\bar{\sigma}}\right)
    \left(1-n_{i+1,\sigma}\right)\rangle,
\end{equation}
where $\sigma=\{\uparrow,\downarrow\}$, $\bar{\sigma}$ are opposite spins 
and $i,i+1$ adjacent sites. This probe determines the probability of 
finding singly occupied neighboring sites with opposite spins.

\begin{figure}[b]
\includegraphics[width=1\columnwidth,clip=true]{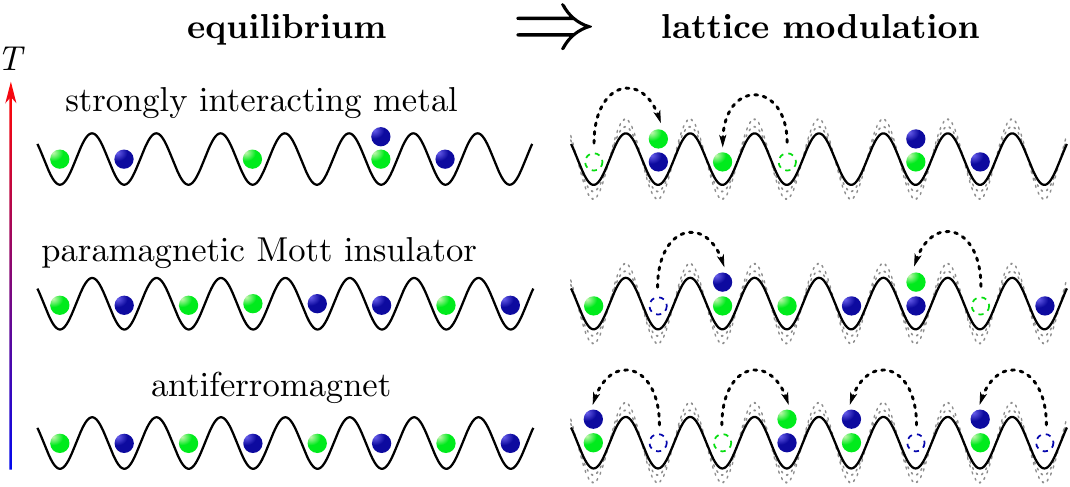}
\caption{(color online) Probing nearest-neighbor correlations for different
phases. A periodic lattice modulation causes tunneling of particles to
neighboring sites. The number of created doublons strongly depends on the
state of the many-body system (strongly interacting metal, paramagnetic
Mott insulator or antiferromagnet) and can be used to determine the
nearest-neighbor correlator $\mathcal{P}_{i,i+1}$ \cite{Kollath2006}.}
\label{fig:cartoon}
\end{figure}

The experimental strategy for detecting the correlator
$\mathcal{P}_{i,i+1}$ relies on exciting the system by a periodic
modulation of the lattice depth \cite{Stoferle2004}. The corresponding
modulation in kinetic energy leads to tunneling of particles to adjacent
sites. If two particles of opposite spin are located on neighboring sites,
additional double occupancies (doublons) are created as shown in Fig.
\ref{fig:cartoon}. The resulting doublon production rate is sensitive on
the nearest-neighbor density and spin correlator $\mathcal{P}_{i,i+1}$. In
the following we describe and characterize this experimental method and
show that in the perturbative regime the frequency integrated doublon
production rate is given by $\mathcal{P}_{i,i+1}$ \cite{Kollath2006,
Huber2009, Sensarma2009, Hassler2009}. We then use this technique to
measure nearest-neighbor correlations as a function of temperature covering
the regime from a paramagnetic Mott insulator to a strongly interacting
metallic state. The results are in good agreement with the predictions of
an \emph{ab initio} theory without any fitting parameters.

The experimental sequence used to produce a quantum degenerate Fermi gas
has been described in detail in previous work \cite{Strohmaier2010,
Jordens2010}. In brief, a balanced spin mixture of $^{40}\text{K}$ atoms in
the $m_F=-9/2$ and $-5/2$ magnetic sublevels of the $F=9/2$ hyperfine
manifold is evaporatively cooled in an optical dipole trap. For samples of
$N=80(7)\times10^3$ atoms we reach temperatures as low as $14\%$ of the
Fermi temperature $T_{\mathrm{F}}$. Subsequently we ramp up a
three-dimensional optical lattice of simple cubic geometry and lattice
constant $d=532\,\mathrm{nm}$. The lattice depth is increased in
$0.2\,\mathrm{s}$ to final values of $7.0(7)\,E_{\mathrm{R}}$ or
$10(1)\,E_{\mathrm{R}}$, where $E_{\mathrm{R}}={h^2}/{8m d^2}$ is the
recoil energy, $h$ is Planck's constant and $m$ denotes the mass of
$^{40}\text{K}$. The hopping $t$ is inferred from Wannier functions
\cite{Jaksch1998} and the on-site interaction energy $U$ is obtained from
lattice modulation spectroscopy \cite{Jordens2008}. The underlying trapping
potential has a mean frequency of $\omega/2\pi=70.1(5)\,\text{Hz}$ for
$7\,E_{\mathrm{R}}$ and $\omega/2\pi=77.3(7)\,\text{Hz}$ for
$10\,E_{\mathrm{R}}$. With this procedure we create samples where the core
is in the Mott insulating regime \cite{Jordens2010}.

After this preparation, the lattice depth is modulated by $\delta V$ along
all three axes in time $\tau$ according to $V(\tau)=V+\delta V \sin(2\pi\nu
\tau)$. This results in a modulation of both the hopping and the on-site
interaction with amplitudes $\delta t$ and $\delta U$ respectively,
creating additional doubly occupied sites as compared to the initial state.
The increase in the number of doublons is maximal when the modulation
frequency $\nu$ coincides with the doublon energy (resonant excitation at
$\nu=U/h$). After the lattice modulation, the fraction of atoms on doubly
occupied sites $D=2 \sum_i \langle n_{i,\uparrow} n_{i,\downarrow}\rangle
/N$ is measured by mapping doublons into a different spin state, subsequent
Stern-Gerlach separation and absorption imaging \cite{Jordens2008}.

\begin{figure}[b]
\includegraphics[width=\columnwidth,clip=true]{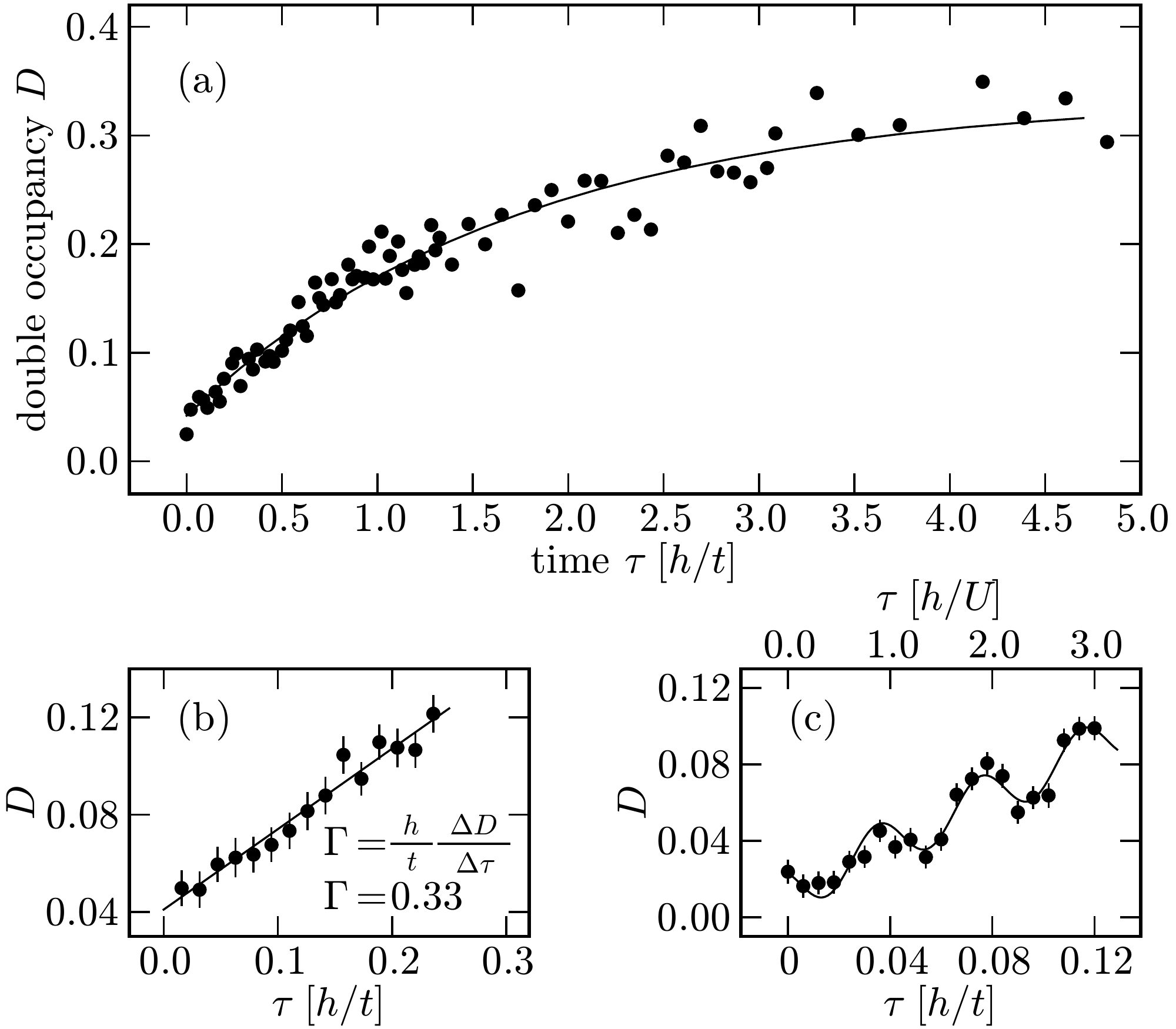}
\caption{Evolution of double occupancy as a function of the lattice
modulation time $\tau$ for resonant excitation. The lattice depth is set to
$7\,E_{\mathrm{R}}$ ($U/6t=4.1$) and the modulation strength is $\delta
V/V=0.1$. (a) The induced double occupancy saturates for large times, which
is well captured by an exponential fit (solid line). (b) At low modulation
times $D$ increases linearly, from which the doublon production rate 
$\Gamma$ is obtained by a linear fit. The lattice depth was set to 
$10\,E_{\mathrm{R}}$ for this measurement ($U/6t=10.6$). (c) On the 
timescale of the modulation period $h/U$ the double occupancy shows an 
underlying low amplitude sinusoidal modulation. The solid line is a fit 
with fixed frequency $U/h$. For clarity, $\delta V/V$ was increased to 
$0.2$. Error bars in double occupancy denote statistical errors from 
multiple measurements.} \label{fig:samplebuilup}
\end{figure}

Fig. \ref{fig:samplebuilup} (a) shows the evolution of double occupancy as
a function of lattice modulation time for resonant excitation. After a
steep initial rise, we observe a saturation of the induced double occupancy
on a timescale on the order of the tunneling time $h/t$. The saturation
value depends only weakly on the modulation strength and reaches typical
values of $20-30 \%$. In contrast to previous work, where only this
saturation regime was considered \cite{Jordens2008, Strohmaier2010}, our
high accuracy in the determination of double occupancy allows us to perform
measurements in the weak excitation limit. Here we find that double
occupancy increases linearly with time as shown in
Fig.~\ref{fig:samplebuilup}(b). We extract the normalized doublon
production rate $\Gamma$ from the slope $\Delta D/\Delta \tau$ of a linear
fit to the data
\begin{equation}
\label{bldprate-def}
\Gamma=\frac{h}{t}\frac{\Delta D}{\Delta \tau}.
\end{equation}
On shorter timescales an underlying oscillatory response at the modulation 
frequency $\nu$ is observed, Fig.~\ref{fig:samplebuilup}(c).

The experiment can be well understood in the framework of time-dependent
perturbation theory, which we outline below. The main result is
that the frequency integrated doublon production rate is proportional to
the nearest-neighbor correlator $\mathcal{P}_{i,i+1}$. The Hamiltonian of
the system can be written as $H=H_0+H_{\mathrm{pert}}(\tau)$, where
$H_0=-tH_t+U H_U$ is the Fermi-Hubbard Hamiltonian and the time-dependent
perturbation is given by $H_{\mathrm{pert}}(\tau)=\left(\delta t
H_{\mathrm{t}} +\delta U H_{\mathrm{U}}\right) \sin(2\pi\nu \tau)$.
Assuming that the doublon production rate is equivalent to the total energy
absorption rate (which becomes exact at half-filling and $U\gg t$), the
perturbation in $U$ can be mapped to an increased tunneling perturbation
with amplitude $\widetilde{\delta t}/t = \delta t/t -\delta U/U$
\cite{Reischl2005}. The response of the system to second order in
perturbation theory is then given by
\begin{eqnarray}
    \label{eq:response}
    D(\tau)&=&D(0)+\left(\frac{\widetilde{\delta t}}{t}\right)
    \chi^{(1)}(\nu) \sin(2\pi\nu \tau)\nonumber \\
    &+&\left(\frac{\widetilde{\delta t}}{t}\right)^2
    \left[\chi^{(2)}(\nu )\frac{\tau}{h/t}+\mathrm{osc.\,terms}\right],
\end{eqnarray}
in good agreement with the experimental observations of Fig.
\ref{fig:samplebuilup} (b),(c) using $\Gamma=
\chi^{(2)}(\nu)(\widetilde{\delta t}/t)^2$. The susceptibilities
$\chi^{(1)}(\nu)$ and $\chi^{(2)}(\nu)$ correspond to the linear response
and the non-oscillatory part of the quadratic response respectively. This
is recovered by Fermi's Golden Rule
\begin{equation*}
    \label{eq:susceptibilites}
    \chi^{(2)}(\nu)=\frac{2\pi^2t}{Nh}\sum_{n}\langle
    n|\delta H_U|n\rangle
\left|\langle n|H_t|0\rangle\right|^2 \delta(\nu-\nu_{n0}).
\end{equation*}
Here $|0\rangle$ and $|n\rangle$ denote the unperturbed and excited states
of $H_0$, $h\nu_{n0}$ is their energy difference and $\delta H_U$ counts
the number of additionally created doublons \cite{Huber2009}. Evaluating
all contributing matrix elements in the limit $U\gg t$, the sum over all
excited states is then equal to $\mathcal{P}_{i,i+1}$. To compare with
experiments we consider the normalized frequency integrated response
\begin{equation}
\label{eq:area}
R=\frac{h}{U}\int\mathrm{d}\nu \Gamma(\nu)=
2\pi^2\left(\frac{\widetilde{\delta
t}}{t}\right)^2 z\frac{t}{U}\mathcal{P},
\end{equation}
where $\mathcal{P}=\sum_{i}\mathcal{P}_{i,i+1}/N$ is the system averaged
correlator and $z$ the connectivity of the lattice. In the
perturbative regime $R$ therefore gives direct access to the
nearest-neighbor correlator $\mathcal{P}$ of the unperturbed initial state.

\begin{figure}[t]
\includegraphics[width=1\columnwidth,clip=true]{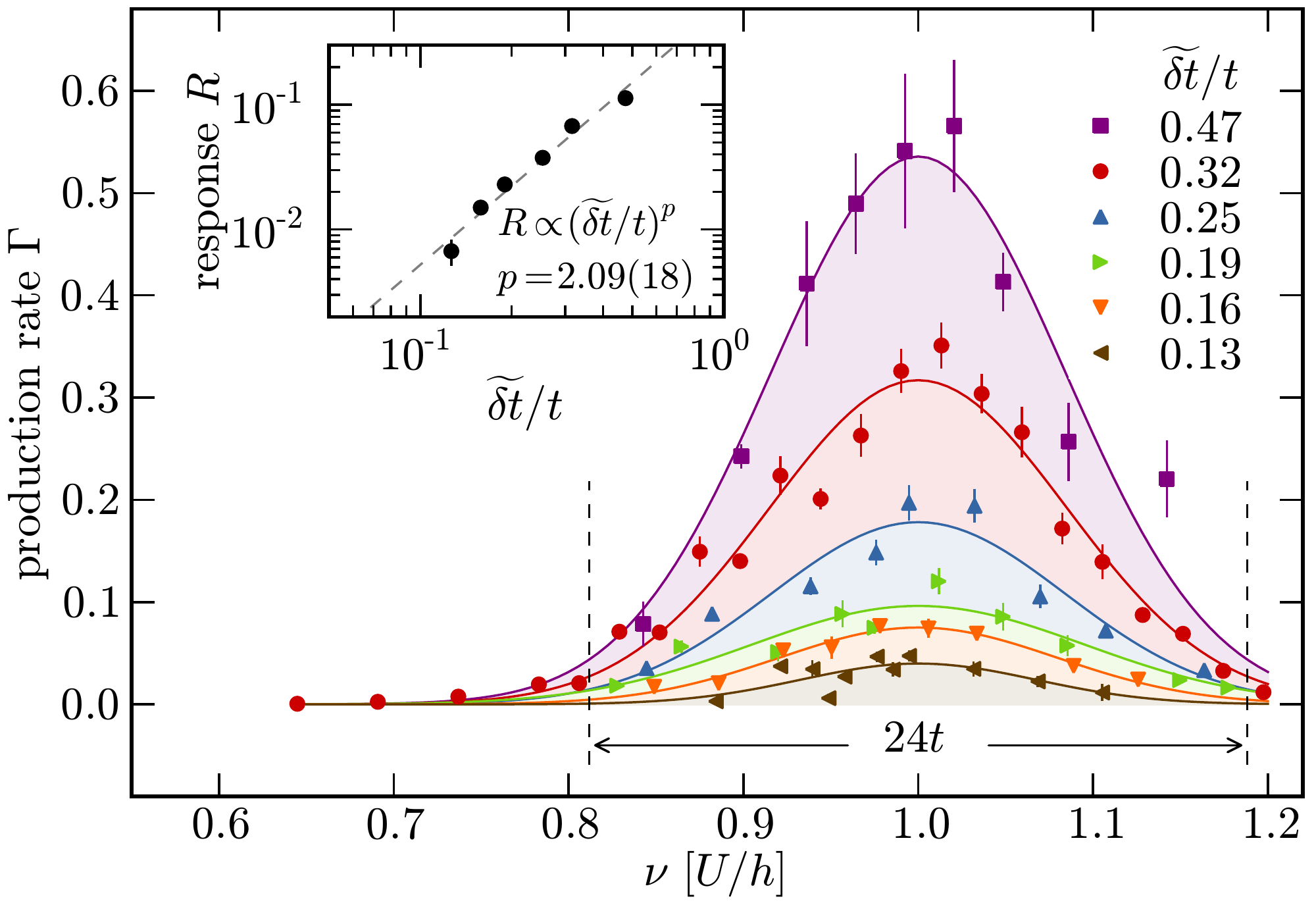}
\caption{(color online) Doublon production rate $\Gamma$ as a function of
the lattice modulation frequency $\nu$, measured for different modulation
amplitudes $\widetilde{\delta t}/t$. The experiments are performed at
$U/6t=10.6$ and $V=10E_{\mathrm{R}}$. The shaded areas are gaussian fits
to the spectra, which are used to extract the frequency integrated response
$R$. The two vertical dashed lines denote twice the three-dimensional
bandwidth $4zt$ (with $z=6$). The inset is a double-logarithmic plot of $R$
for various modulation amplitudes, where the dashed line is a power law
fit. Error bars denote the fit errors.}
\label{fig:scaling}
\end{figure}

We validate that the experiments are performed in the weak excitation
regime by studying the scaling of $R$ with the relative modulation
amplitude $\widetilde{\delta t}/t$ \footnote{We use $\widetilde{\delta
t}/t= -\sqrt{V/E_{\mathrm{R}}}\delta V/V$, which is accurate to a few
percent for the lattice depths used in the experiment.}.
We measure the doublon production rate as a function of the modulation
frequency, from which the modulation spectra in Fig.
\ref{fig:scaling} are obtained. The frequency integrated response is then
determined by a gaussian fit to each spectrum. The result is plotted as a
function of the lattice modulation amplitude $\widetilde{\delta t}/t$
(inset of Fig. \ref{fig:scaling}). We find a scaling exponent of
$2.09(18)$, in very good agreement with the expected value of $2$
predicted by second order perturbation theory. We can thus infer the
nearest-neighbor correlator $\mathcal{P}$ from the frequency integrated
response.

Further information can be obtained from the lineshape of the modulation
spectra, which reveals the density of states of the excitations. At half
filling and temperatures well above the N\'eel transition, the density of
states has an approximately triangular shape of full width $\sim 3 z t$
\cite{Sensarma2009}. However, the trapping potential is expected to
broaden the spectrum and introduce deviations to the lineshape
\cite{Kollath2006}. This is consistent with the experimental data, which
is well captured by gaussian fits with $1/e^2$ diameters of $4zt$. The
doublon production rate on resonance shows the same scaling behavior as
the integrated response $R$, with a scaling exponent of $1.98(11)$. This
allows us to determine $\mathcal{P}$ from the resonant doublon production
rate alone assuming a gaussian density of states as in Fig.
\ref{fig:scaling} \footnote{We have verified this assumption by inferring
the value of the nearest-neighbor correlator from the inset of Fig.
\ref{fig:scaling}, which is in good agreement with the model.}.

We now use doublon production rate measurements to determine the
nearest-neighbor correlator $\mathcal{P}$ as a function of entropy for
$U/6t=4.1$. We prepare samples with different entropies per particle by
adding a variable waiting time in the optical dipole trap of up to
$2\,\mathrm{s}$, which results in heating due to inelastic scattering
processes. The entropy per particle $s_{\mathrm{in}}$ before loading into
the lattice is inferred from Fermi fits to the momentum distribution of the
cloud after expansion. This is a lower bound for the specific entropy in
the lattice, as non-adiabatic processes take place during the loading. An
upper bound $s_{\mathrm{out}}$ is given by the entropy measured after
reversing the loading procedure.

\begin{figure}[t]
\includegraphics[width=1\columnwidth,clip=true]{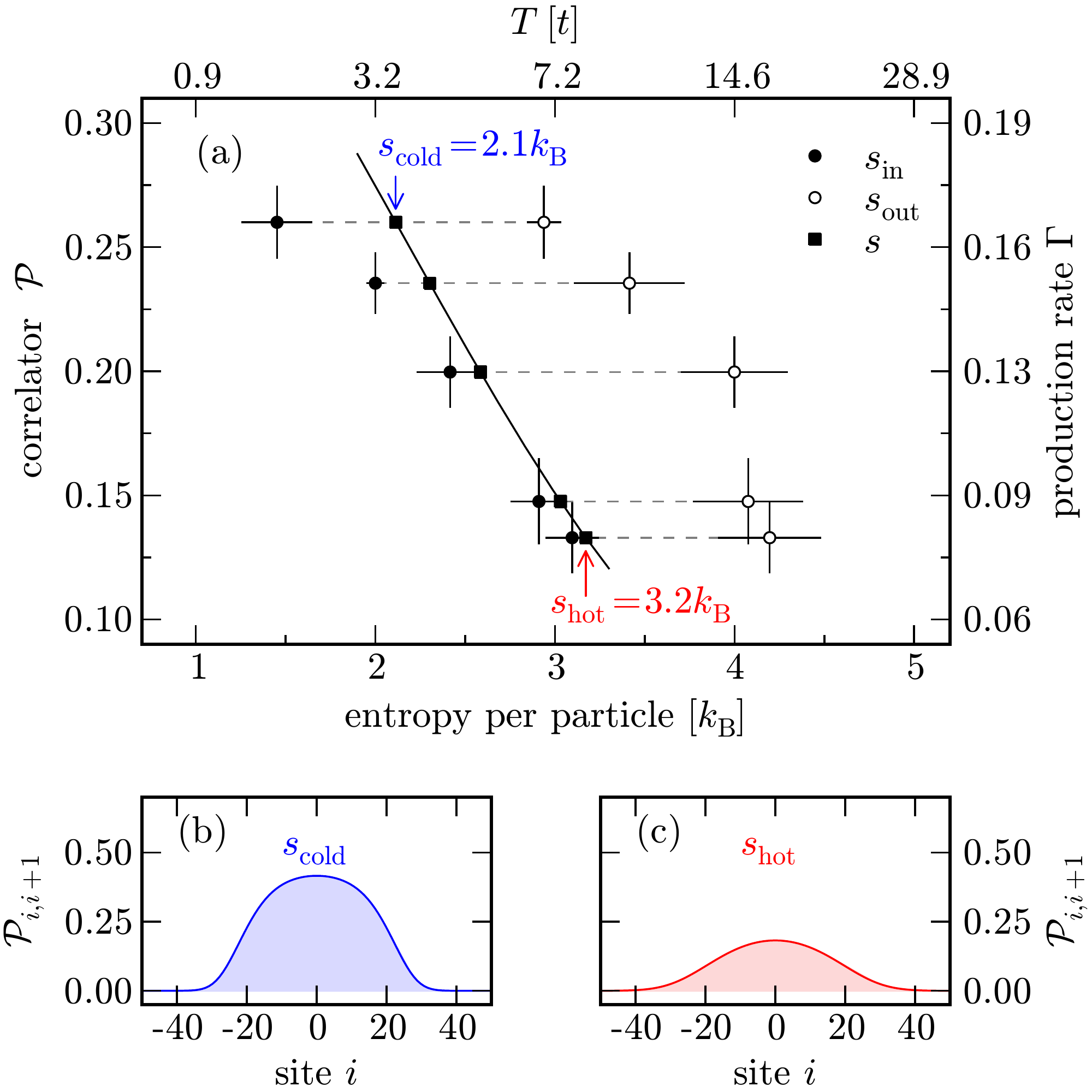}
\caption{(color online) Measurement of the nearest-neighbor correlator
$\mathcal{P}$. (a) Dependence on entropy per particle in a trapped lattice
system for $U/6t=4.1$, $V=7E_{\mathrm{R}}$ and $\widetilde{\delta
t}/t=0.26$. Solid and open circles denote lower and upper bounds
$s_{\textrm{in}}$ and $s_{\textrm{out}}$ for the entropy per particle in
the lattice for each measured value of $\mathcal{P}$. The black solid line
is the calculated nearest-neighbor correlator obtained from second order
HTSE without any fitting parameters. The entropy $s$ inferred from
comparing each measured value of $\mathcal{P}$ to theory lies in between
the experimental bounds. For clarity the correspoding lattice temperatures
$T$ and measured doublon production rates $\Gamma$ are also included.
Error bars denote statistical errors from several measurements. (b) and
(c) show the calculated distribution of $\mathcal{P}_{i,i+1}$ in the trap
for the lowest and highest entropies $s_{\mathrm{cold}}$ and
$s_{\mathrm{hot}}$, while the area under these curves corresponds to the
measured correlator $\mathcal{P}$.} \label{fig:correlator-demo}
\end{figure}

The nearest-neighbor correlator rapidly decreases with increasing entropy,
as shown in Fig. \ref{fig:correlator-demo}(a). This behavior has a simple
physical interpretation: in a harmonically trapped lattice system higher
temperatures lead to an increased cloud size, which results in a large
number of empty sites. The probability of finding two neighboring singly
occupied sites is therefore strongly reduced. This qualitative picture is
confirmed by \emph{ab initio} calculations of the nearest-neighbor
correlator. We use a high temperature series expansion (HTSE) up to second
order in $t/k_{\mathrm{B}}T$, where $k_{\mathrm{B}}$ is the Boltzmann
constant \cite{tenHaaf1992, Henderson1992}. Due to the harmonic confinement,
the trap averaged correlator $\mathcal{P}$ needs to be evaluated. This is
done using a local density approximation, which is an excellent assumption
in this temperature regime \cite{Scarola2009}. The system parameters are
calibrated by independent methods \cite{Jordens2010}. The results of this
calculation are shown in Fig. \ref{fig:correlator-demo}(a).

We find quantitative agreement between the measured nearest-neighbor
correlator and the theoretical predictions without any fitting parameters.
The theoretical entropies $s$ corresponding to the measured values of
$\mathcal{P}$ lie in between the lower and upper experimental bounds
$s_{\textrm{in}}$ and $s_{\textrm{out}}$. Nearest-neighbor correlation
measurements thus allow us to determine entropy and temperature in the
lattice for regimes where theory is still reliable and a quantitative
comparison is possible. In contrast to thermometry in the optical dipole
trap, the correlator $\mathcal{P}$ is a direct observable in the lattice
and does not rely on adiabaticity assumptions during the loading
procedure. A comparison of these two methods suggests increased heating
during the loading of the lattice for colder initial temperatures.

Further insight can be obtained from the theoretical model by
investigating the spatial distribution of the nearest-neighbor correlator
over the trap. From the inferred entropies $s$ we calculate the profiles
shown in Fig. \ref{fig:correlator-demo}(b) and (c) for the parameters of
our system \footnote{The local value of the nearest-neighbor correlator is
obtained by combining the HTSE result with a local density approximation.
The average over all sites then yields the values of $\mathcal{P}$ shown
in Fig. 4(a).}. At half filling and deep in the Mott insulating regime
$\mathcal{P}_{i,i+1}$ is expected to be close to $0.5$, whereas thermal
excitations reduce this number in the metallic phase. This is confirmed by
the values in the core of the system, with $0.44$ for the coldest and
$0.17$ for the hottest point. The decrease in $\mathcal{P}_{i,i+1}$ thus
signals the transition from a paramagnetic Mott insulator to a strongly
interacting metal. As the onset of local spin correlations corresponds to
an increase of $\mathcal{P}_{i,i+1}$ above $0.5$, nearest-neighbor
correlation measurements are a promising tool for studying the approach to
the antiferromagnetic phase \cite{Fuchs2010}.

In conclusion, we have measured nearest-neighbor correlations of ultracold
fermions in optical lattices by determining the response of the system to a
weak lattice modulation. This observable is well suited for thermometry in
the lattice and can be used to explore novel cooling schemes
\cite{Bernier2009}. The technique opens new prospects for studying the
approach to the antiferromagnetic phase, since the regime between a
paramagnetic Mott insulator and an antiferromagnet is governed by the
formation of short-range magnetic correlations. In the future,
nearest-neighbor correlation measurements might give insight into
resonating valence bond ground states, where singlet correlations on
neighboring sites are expected to occur in the absence of long range
ordering \cite{Anderson1987}.

We are grateful to N.~Strohmaier and H.~Moritz for contributions in the
early stages of the experiment, and to G.~Jotzu for a critical reading of
the manuscript. We thank M.~A.~Cazalilla, E.~Demler, T.~Giamarchi,
F.~Hassler, A.~Ho, S.~Huber, C.~Kollath, D.~Pekker, L.~Pollet,
A.~R\"{u}egg, R.~Sensarma and A.~Tokuno for insightful discussions, and
SNF, NCCR-MaNEP, NAME-QUAM (EU, FET open) and SQMS (ERC advanced grant)
for funding.

\end{document}